\documentstyle[epsfig,a4]{article}

\advance\hoffset by -1.7 cm
\begin{document}

\begin{tabbing}
\hskip 14 cm \= {MRAO-}\\
\> DAMTP-96-61\\
\> Imperial/95/96/73\\
\> Submitted to {\em PRD}\\
\> October 1996 \\
\end{tabbing}

\openup1\jot

\begin{center}
{\Huge\bf Ideal scales for weighing the Universe}
\vskip 1.2cm
{\large\bf Jo\~ao Magueijo$^{(1)}$ and M.P.~Hobson$^{(2)}$}\\
$^{(1)}$ The Blackett Laboratory, Imperial College,
Prince Consort Road, London SW7 2BZ, UK\\
$^{(2)}$ Mullard Radio Astronomy Observatory,
Cavendish Laboratory\\ Madingley Road, Cambridge, CB3 0HE, UK.\\
\end{center}

\begin{flushleft}
\end{flushleft}

\begin{abstract}
\openup1\jot
We investigate the performance of a large class of
cosmic microwave background experiments with respect to their ability
to measure various cosmological parameters. We pay special attention
to the measurement of the total cosmological density,  
$\Omega$. We consider interferometer experiments, all-sky single-dish
experiments, and also single-dish experiments with a deep-patch technique.
Power spectrum estimates for these experiments are studied, and
their induced errors in cosmological parameter estimates evaluated.
Given this motivation we find various promising corners in the experiment 
parameter space surveyed. Low noise all-sky
satellite experiments are the expensive option, but they are best suited
for dealing with large sets of cosmological parameters. 
At intermediate noises
we find a useful corner in high-resolution deep patch single-dish experiments.
Interferometers are limited by sample variance, but provide the best estimates
based on the very small angular scales.
For all these experiments we present conservative, but still promising
estimates of the accuracy of the measurement of $\Omega$. In these estimates
we consider a variety of possible signals not necessarily in the vicinity
of  standard cold dark matter.
\end{abstract}

\thispagestyle{empty}
\pagebreak
\setcounter {page}{1}

\section{Introduction}

In many recent papers \cite{be,st,kn,j,j1} it has been shown how
the cosmic microwave background (CMBR) angular power spectrum 
$C_\ell$ may  provide a clean 
measurement of several cosmological parameters, and also of the so-called
inflationary observables. In particular the Universe total density in units
of the critical density, $\Omega$, is known to leave a distinct 
imprint in the $C_\ell$ spectrum \cite{hs}. 
Therefore it has been suggested \cite{j,j1}
that the next generation of experiments, 
in particular MAP \cite{map} and COBRAS/SAMBA \cite{cobras/samba}, should
settle the question of the total density of the Universe. 
This would also determine the geometry of the Universe, 
deciding at long last which of the three types
of Freedman models (spherical, planar, hyperbolic) 
describes our Universe.
A standard error bar in $\Omega$ of about 10\% is expected 
for an experiment with features
similar to what is now known as the MAP proposal \cite{map}. 
In this estimate one marginalizes with respect to a large parameter
space, with unassuming priors. 

In this paper we reexamine and extend the work in \cite{j,j1}
where estimates for $\sigma(\Omega)$ were computed for
prototype satellite experiments. 
We briefly assess the errors and uncertainties in the covariance 
matrix approach used in \cite{j,j1}.
We find that the results obtained by this method are to be seen as a guide
only. Slight changes in the algorithm, as well as in the number of parameters
considered, may easily raise a standard error in $\Omega$ from,
say 10\%, to about
30\%. Also in the analysis performed in \cite{j,j1} it was assumed that
the ``signal'' comes from a standard cold dark matter theory (sCDM).
We find the unsurprising result that if one considers instead the case of
signals coming from theories with a scalar tilt smaller than 1, 
or low $\Omega$
models, then the errors are significantly larger. Since we don't a priori
know what the signal is, one must be prepared for the worse when
estimating the performance of a given experiment. Overall we find that
the analysis in \cite{j,j1} is a perfectly valid preliminary
estimate. However, when reinserted in the context of its own uncertainties,
and reapplied to less generous signals, this analysis reveals that the
conclusions in \cite{j,j1} are perhaps embedded in what might be
described as realistic optimism. In this paper we take the 
opposite point of view: 
we decide to be pessimistic, not to be cumbersome, but instead
in order to investigate how one could vary  the observational 
strategy in order to improve final estimates of $\Omega$.

We consider a covariance matrix  set up yielding the worst prediction for
sCDM, and also consider the case of signals coming from non-sCDM theories
which make $\Omega$ more difficult to measure. However we also generalize
considerably the space of possible experiments covered by the analysis
in \cite{j,j1}. Taking up work in \cite{us,us1}
we consider single-dish experiments in which the sky-coverage
is deliberately small (the so-called deep patch technique). We also 
consider the case of interferometric experiments (see \cite{readhead}
for a good review).  We then
try to design the ideal experiment for measuring $\Omega$,
given a constraint mathematically rephrasing fixed finite funding.  
This work provides guidance along two lines. Firstly, one may 
examine what is the ideal scanning method for best results  
subject to the constraint. Secondly, one may provide the value
of the error in $\Omega$ as a function of this constraint, assuming
ideal scanning. This will set up lower bounds on the constraint
for a meaningful experiment, telling us also how fast
these errors will go down thereon, after a given constraint improvement.
Not surprisingly we find more than one promising corner in the experiment
parameter space surveyed. In summary, all-sky low noise satellite experiments,
high resolution intermediate noise ground based experiments, and 
interferometers, all, for different reasons, provide estimates for
$\Omega$ with useful errorbars.  It is now up to political details 
to decide who will win the race of the weighing of the Universe.

Our paper is organized as follows. In Section~\ref{review} we review
the covariance matrix approach to the estimation of cosmological
parameters. However we tie this approach to previous work we have done 
\cite{us1} on power spectrum estimates for a large class 
of experiments. Then in Section~\ref{stat} we make use of this 
formalism to produce heuristic arguments on what are the best scales 
for measuring various  cosmological parameters, with special 
reference to $\Omega$. We then show how the scanning strategy may 
improve experimental performance on sections of the spectrum which are
most relevant for the determination of $\Omega$. We show cases where
reducing sky-coverage is useful for improving the measurement
of $\Omega$, and cases where it is not. 
Given these insights we then jump into the full problem of estimating
the errorbar in $\Omega$ given the specifications of an experiment
ideally suited to measure $\Omega$ subject to its constraints.
In Section~\ref{cdm} we present results for signals coming from the vicinity
of sCDM, and in Section~\ref{ncdm} we consider other signals.  
In Section~\ref{fewpar}
we also spell out formalism uncertainties and how they affect 
the results in  Section~\ref{fewpar}.  As a side remark 
in Section~\ref{intsagain} we 
investigate the interferometers' performance with respect to 
other cosmological parameters as well. We conclude with a few comments
on what steps should be taken towards a more precise and
comprehensive analysis of determination of cosmological parameters.

\section{Determination of cosmological parameters with CMB experiments}
\label{review}

We start by reviewing the covariance matrix approach \cite{j1,ag,etc}
in the guise to be used in this paper. To set the notation,
the $C_\ell$ spectrum is  defined
from the two-point correlation function $C(\theta)$ as
\begin{equation}
C(\theta)= \sum_{\ell=2}^\infty {2\ell+1\over 4\pi} C_\ell
P_\ell(\cos\theta),
\end{equation}
or alternatively from $C_{\ell} = \langle |a_{\ell
m}|^2\rangle$, where the $a_{\ell m}$ are the spherical harmonic
coefficients in the expansion
\begin{equation}
\frac{\Delta T(\bf {\hat{x}})}{T}   
= \sum_{\ell=0}^{\infty} \sum_{m=-\ell}^{\ell}
a_{\ell m} Y_{\ell m} (\bf {\hat{x}}).
\end{equation}
The $C_\ell$ spectra will here be used to estimate a maximal parameter space
spanned by ${\bf s}=\{Q,\Omega,$ $\Omega_b h^2,$ $h,n_S,\Lambda,$
$r,n_T,\tau,\Omega_{\nu}\}$. Here $\Omega$, $\Omega_b$, and
$\Omega_\nu$ are the total, the  baryon, and the heavy neutrino densities, in
units of the critical density. When $\Omega_\nu\neq 0$ we have 
assumed one massive neutrino. $h$ parameterizes the Hubble constant
in the usual way: $H=100 h {\rm Km }\;{\rm sec}^{-1} {\rm Mpc}^{-1}$. 
$\Lambda$ is the cosmological constant. Whenever $\Lambda\neq 0$
we have defined the total density of the Universe, responsible for
its geometry, to be $\Omega=\Omega_0+\Lambda$, where $\Omega_0$
is the total density in matter, radiation, and neutrinos.
$n_S$ and $n_T$ are the tilts in the primordial
power spectrum of scalar and tensor perturbations, respectively.
$Q$ is the quadrupole in $\mu K$, so that $C_2=(4\pi/5)(Q/T_0)^2$,
where $T_0$ is the average CMBR temperature. We shall normalize all spectra 
by the $10^\circ$ rms temperature. $r$ is the ratio of scalar and tensor
components defined at $\ell=2$, that is $r=C^T_2 /C^S_2$.
Finally $\tau$ is the optical depth to the epoch of recombination.
Besides this maximal set of parameters we will also consider subsets
in order to emphasise the dependence of the final results on the
number of uncertain parameters. 

We shall use a Gaussian approximation to the likelihood function
as proposed in \cite{j1,ag,etc}.
This consists of approximating the likelihood linearly near its
maximum, resulting in 
\begin{equation}
{\cal L}({\bf s})\propto \exp \left[ -{1\over 2}({\bf s}-{\bf s}_0)
\cdot [\alpha] \cdot({\bf s}-{\bf s}_0)\right],
\end{equation}
where ${\bf s}_0$ is the underlying theory set of parameters, and
the curvature matrix $[\alpha]$ is given by
\begin{equation}
\alpha_{ij} = \sum_\ell {1\over\sigma^2(C_\ell)}
\left[{\partial C_\ell({\bf s}_0)\over\partial s_i}
{\partial C_\ell({\bf s}_0)\over\partial s_j}\right].
\end{equation}
Inverting $[\alpha]$ one therefore obtains the parameters covariance
matrix $[{\cal C}]=[\alpha]^{-1}$, thereby converting the errors
in the measurement of the $C_\ell$ (given by $\sigma^2(C_\ell)$)
into errors in the estimates of ${\bf s}$. The diagonal components,
${\cal C}_{ii}$, in particular, are the errors $\sigma^2(s_i)$ in
the parameter $s_i$ obtained by marginalizing with respect to all 
other parameters. We will find that the marginal errors $\sigma^2(s_i)$
depend very strongly on the total set of parameters chosen.

The errors in the estimates of the power spectrum $\sigma(C_\ell)$ 
are due to cosmic/sample variance, instrumental noise, and 
foreground subtraction.
Following \cite{readhead,scott} we shall translate the effects of foreground 
subtraction into a renormalized noise term, that is we take foreground
subtraction errors into account by multiplying the instrumental noise by a 
deterioration factor. The deterioration factor is a priori unknown.
An optimist would place it between 1 and 2, a pessimist above 3. 
We have empirically found this procedure to be appropriate for
interferometer experiments \cite{cat} by directly examining the effects
of deconvolving galactic foregrounds in  real data using maximum likelihood
methods. We shall consider errors in $C_\ell$ 
estimation for a large class of experiments. These include 
all-sky single-dish experiments, single-dish
experiments with a deep-patch technique, and interferometers.
Power spectrum estimate errors for these experiments were studied
in \cite{us,us1},  where the reader may find more detail and derivations.
Here we merely review the main results in \cite{us,us1}. 

\subsection{A satellite experiment}

We first review the case of a single-dish
experiment covering a large portion of the sky $f_{\rm sky}$
(see \cite{kn,j1,us1} for more details). Let the 
pixel area be $\Omega_{\rm pix}=\pi\theta_b^2$, where $\theta_b
\approx 2.35 \sigma_b$ is the FWHM size of each pixel, assumed to be
Gaussian shaped. The probed spectrum will then appear multiplied by a beam
factor $B(l)=\exp(l^2/2\sigma_b^2)$, which needs to be deconvolved. 
Now let the noise per pixel be $\sigma^2_{\rm pix}=s^2/t_{\rm pix}$,
where $s$ is the detector sensitivity and $t_{\rm pix}$ is the time spent
observing each pixel. Then fixing the detector sensitivity $s$
and total time of observation $t_{\rm tot}$ fixes the quantity
$w^{-1}=4\pi s^2/t_{\rm tot}=\sigma^2_{\rm pix}\Omega_{\rm pix}/f_{sky}$,
where $f_{\rm sky}$ is the fraction of sky coverage.
We therefore parameterize the noise level with $w^{-1}$.
In this argument we have assumed active incomplete sky-coverage, that is,
an experimental strategy in which one deliberately focuses on an 
incomplete patch of the sky. Because this is done deliberately, the 
integration time on each pixel is  larger. Therefore the errors due
to noise are smaller, even though the errors due to sample variance are
larger. This set up is also sometimes called deep-patch technique.
On the other hand, passive incomplete sky-coverage occurs
if one observes the whole sky, but then portions of it have to be thrown out 
for various reasons, eg. point source contamination, or galactic
obscuration. In such a case the noise parameter is
$w^{-1}=4\pi s^2/t_{\rm tot}=\sigma^2_{\rm pix}\Omega_{\rm pix}$.
Then using the results in \cite{us,us1} one may prove that
\begin{equation}
\sigma^2(C_\ell) = {2C_\ell^2\over (2\ell +1)f_{\rm sky}}
\left[1+ {w^{-1} f_{\rm sky}\over 
C_\ell B(l)^2} \right]^2,
\label{variance}
\end{equation}
for active incomplete sky-coverage.
For passive incomplete sky-coverage the appropriate formula is 
\begin{equation}
\sigma^2(C_\ell) = {2C_\ell^2\over (2\ell +1)f_{\rm sky}}
\left[1+ {w^{-1} \over 
C_\ell B(l)^2} \right]^2.
\label{variance1}
\end{equation}

\subsection{A deep-patch single-dish experiment}

These formulae break down if $f_{\rm sky}\ll 1$. The effects of 
small sky coverage have been studied in \cite{us,us1} and can 
be summarized as follows. Firstly the probed spectrum becomes 
the convolution of the raw spectrum with the window spectrum. 
For definiteness let us assume that the field is a square with size $L$, 
in radians. Then the probed spectrum will differ from the raw spectrum
in that features on a scale $\Delta l\approx 1/L$ are smoothed out, 
and also a white noise tail appears for $l<2\pi/L$. If the distortions
introduced are considerable, then a deconvolution recipe becomes 
necessary, increasing the errors significantly. We have however 
checked that sCDM Doppler peaks
do not require deconvolution for any window with $L>4^\circ$,
as long as a bell is applied to the field, if this has edges.
The $l<2\pi/L$ section of the spectrum cannot be
recovered, but apart from this the probed and the raw spectra are
proportional to each other. If $\Omega<1$ the lower
bound on $L$ for avoiding the need for deconvolution becomes even
less restrictive (of order $L>2^\circ$ for $\Omega=0.3$).

Secondly incomplete sky coverage introduces correlations between 
the modes used in the $C_\ell$ estimates. This reduces spectral 
resolution, that is, it allows independent $C_\ell$ estimates only with
a separation $\Delta l>1$ (see \cite{us1} for formulae for $\Delta\ell$). 
Correlations also reduce the number of independent modes available for 
each independent estimate, and therefore cosmic variance is increased
by finite sky-coverage (an effect sometimes called sample variance).
These two effects can be 
studied quantitatively  by stereographically projecting the observation
field onto a plane, Fourier transforming, and  replacing the
Fourier plane by what we called an uncorrelated mesh of modes.
This mesh contains only quasi-uncorrelated modes, and their
density equals the density of ${\it independent}$ modes present
in the full Fourier plane itself.
For a square field with a side $L$ treated with a cosine bell 
\cite{teg} the side of the uncorrelated mesh is $k_0\approx 2\pi/L$. 
From the uncorrelated mesh modes ${\bf k}_i$ one may then extract
power spectrum estimates for values of $l_i$ for which mesh points
exist satisfying $l_i={\rm int}(|{\bf k}_i|)$, where int denotes
the integer part. In general one may come up with
uncorrelated estimates only with a separation 
$\Delta \ell\approx {\sqrt {l^2+k_0^2/\pi}}-\ell$, for $\ell>k_0$.
Only for $\ell>k_0^2/(2\pi)$ can individual $C_\ell$ be estimated 
($\Delta \ell\approx 1$). 
The curvature matrix $[\alpha]$ corresponding
to these power spectrum estimates is then given by
\begin{equation}
\alpha_{ij} = \sum_{\ell_i} {1\over\sigma^2(C_{\ell_i})}
\left[{\partial C_\ell({\bf s}_0)\over\partial s_i}
{\partial C_\ell({\bf s}_0)\over\partial s_j}\right],
\label{curv}
\end{equation}
where the  errors in the estimates $\sigma^2(C_{\ell_i})$ are 
\begin{equation}
\sigma^2(C_{\ell_i})\approx {2C_{\ell_i}^2\over N(\ell_i)}
\left(1+{w^{-1} f_{\rm sky}\over
B^2(\ell_i) C(\ell_i)}\right)^2,
\label{varnoise1}
\end{equation}
where $N(\ell_i)$ is the number of modes satisfying 
$l_i={\rm int}(|{\bf k}_i|)$. 

If there is near all-sky coverage then $\Delta\ell\approx 1$ and
$N(\ell)=(2\ell +1)f_{\rm sky}$, recovering the large sky-coverage
formulae. In fact the uncorrelated mesh results for $L\approx 202^\circ$
(corresponding to $f_{\rm sky}\approx 1$) differ from the large
sky-coverage limit results by less than 1\%. On the other hand for
$L<100^\circ$ (that is for $f_{\rm sky}<0.25$) the errors inferred from
the large sky-coverage limit are always grossly underestimated.
Therefore the uncorrelated mesh formalism acts as a generalization
of the large sky-coverage formulae.

\subsection{Interferometers}

For interferometers the field is Gaussian shaped (the so-called 
primary beam \cite{readhead,cat}). Let $\theta_w=2.35\sigma_w$ 
be the primary beam FWHM. Deconvolution problems may be avoided 
for sCDM by imposing a primary beam size $\theta_w>4^\circ$ 
($\theta_w>2^\circ$ for $\Omega=0.3$).
Interferometers make measurements directly in Fourier 
space (which in the community jargon is called the $uv$-plane). 
Again uncorrelated estimates and their variances  may be  
obtained by means of an uncorrelated-mesh \cite{us1}, 
this time with a side $k_0\approx 2{\sqrt {4\pi\log 2}}/\theta_w$. 
Also if one decides to observe $n_f$ well-separated fields, then
each mesh-point acquires an extra index $i=1,\ldots, n_f$,
and points with different indices are uncorrelated. 
With these modifications the curvature matrix may then be computed 
as in (\ref{curv}).

Now let $N_{\rm vis}$ be the number of visibilities in each uncorrelated
mesh cell, let $s$ be the sensitivity of the detectors, and 
$t_{\rm vis}$ be the time spent observing each visibility. 
The coverage density is given by $\rho_c=N_{\rm vis}t_{\rm vis}\Omega_w/t_f$,
where $t_f$ is the time spent on each field and $\Omega_w=\pi
\sigma_w^2 $ is the field area. We shall assume that the coverage
density is uniform in a ring of the $uv$-plane going from $k_0$ to 
a certain $k_{max}$ delimiting the outermost $uv$-tracks depicted
by the interferometer. This may be attained with a dish geometry 
like the one proposed in \cite{intf}.  

In order to parameterize the noise for interferometers, we
now notice that for fixed detector sensitivity and total observation
time one should now keep constant $w^{-1}=(2\pi)^2s^2/(\rho_c t_{\rm tot})=
(2\pi)^2\sigma^2_N/(\Omega^{s3} n_f)$. This allows comparing experimental
strategies subject to the same constraint (time and money) but choosing,
say, different primary beam sizes and number of fields.
As shown in \cite{us1} the power spectrum estimates for interferometers
provided by their uncorrelated mesh are now affected by the errors:
\begin{equation}
\sigma^2(C_{\ell_i})={2C_{\ell_i}^2\over N(\ell_i)}
\left(1+{w^{-1}\Omega^{s2} n_f\over 
C(\ell_i)}\right)^2.
\label{varnoise2}
\end{equation}
It should be remarked that as we go up  in $\ell$ the signal-to noise
in each independent mode goes down as a power law (we are assuming
that the outer ring is before the Silk damping tail). This is to 
be contrasted with single-dish experiments, where the same
signal-to-noise always goes down like an exponential in $\ell$.  
This feature makes interferometers desirable for the measurement 
of high $\ell$ features. 
\begin{figure}
\centerline{\epsfig{file=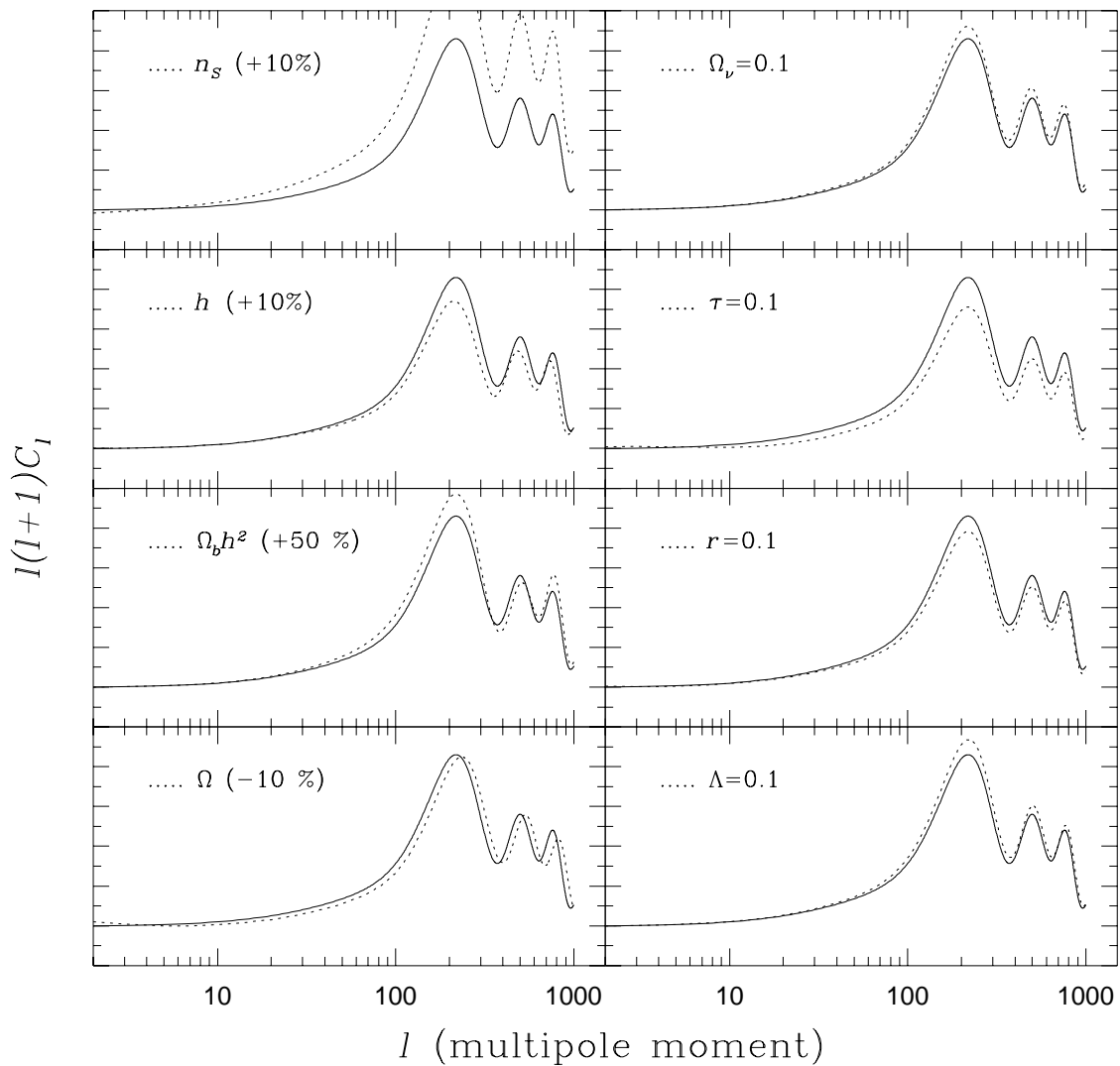,width=15cm}}
\caption{The $C_\ell$ spectrum of sCDM (line) and of theories in its
vicinity obtained by changing one of the parameters $s_i$ to a value
indicated in the label (points). The changes introduced may be mild
or radical depending on the parameter. Also there is some level of
degeneracy in the changes introduced.}
\label{fig1}
\end{figure}

\section{Relevant scales for determining cosmological parameters}
\label{stat}

We now investigate which are the statistically most relevant $C_\ell$'s
for the determination of the various cosmological parameters. To do this
we start by assuming for each parameter $s_i$
that all other parameters are
kept fixed, so that 
\begin{equation}\label{fact}
{1\over \sigma^2(s_i)}=\sum_\ell {S(l; s_i)\over N(\ell)},
\end{equation}
defining a sensitivity factor $S(l;s_i)$
\begin{equation}
S(l;s_i)=\left[\partial C_\ell({\bf s}_0)\over\partial s_i\right]^2
{2l+1\over 2 C_\ell^2},
\end{equation}
and a noise factor $N(\ell)$ which for 
a single-dish experiment with active limited sky coverage
is given by
\begin{equation}
N(\ell)=f_{\rm sky}^{-1}
\left[1+ {w^{-1} f_{\rm sky}\over 
C_\ell B(l)^2} \right]^2
\end{equation}
The sensitivity factor $S(l;s_i)$ tells us the relevance of each
$C_\ell$ for the determination of the parameter $s_i$ subject only
to the unremovable cosmic variance. 
The noise factor $N(\ell)$ will then deteriorate
this result. Knowledge of where $S(l;s_i)$ is largest will then 
suggest experimental features. If one is most interested in a given
parameter $s_i$ then one should require that $N(\ell)$ be the closest
to $1$ (its mathematical minimum)
where $S(l;s_i)$ is the largest. In practice the determination
of a given $s_i$ is coupled to the determination of all the other
parameters. Still one may detect general trends by studying
the $S(l;s_i)$ curves.
\begin{figure}[t]
\centerline{\epsfig{file=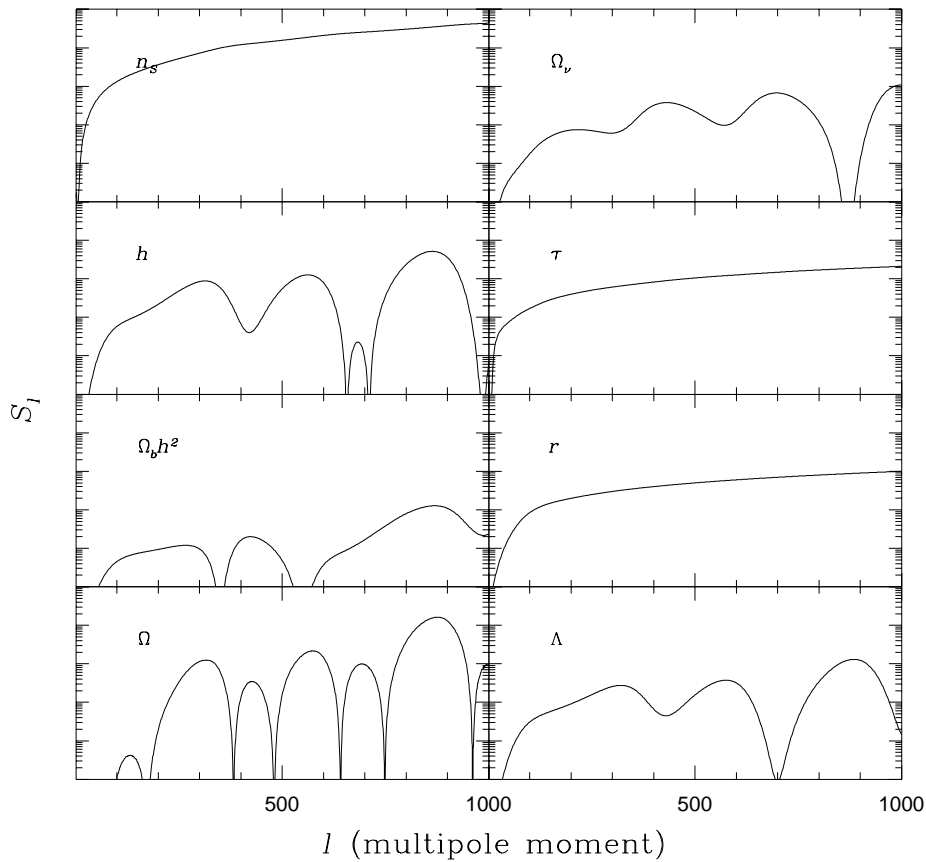,width=12cm}}
\caption{The sensitivity functions $S(l;s_i)$ for various cosmological 
parameters as inferred from the variations used in Fig.1. 
We have taken percentage sensitivities $S(l;s_i)s_i^2$
on the plots on the left hand column.}
\label{fig2}
\end{figure}

\subsection{The sensitivity to cosmological parameters}

The $C_\ell$ spectrum dependence on ${\bf s}$ is illustrated in 
Figure~\ref{fig1}. We have assumed that we are in the vicinity of sCDM 
defined as ${\bf s}=(18,1,0.01,0.5,1,0,0,0,0,0)$. We have then shown
spectra obtained by changing various parameters $s_i$ 
keeping all other parameters fixed at their sCDM value. 

Two comments should be made.
Firstly the function $C_\ell({\bf s})$ changes at rather different
rates depending on the parameter $s_i$. The scalar tilt $n_S$
and the total cosmological density $\Omega$ change the whole
spectrum rather drastically when subject to
small percentage variations. A multiplicative shift
in $l$ is the benchmark of $\Omega$. An overall tilt of the spectrum
is of course the hallmark of $n_S$.
Late reionization ($\tau\neq 0$) and tensor modes ($r\neq 0$)
affect strongly the relative
normalizations of the plateau and peaks. The separate
shapes of these two spectrum sections, however, remain roughly unchanged. 
Hence a low coverage high resolution experiment, and a high
coverage low resolution experiment, considered on their own
would effectively be blind to $\tau$ and $r$.
The other parameters considered leave rather subtle imprints.
The baryon content enhances odd peaks relative to even peaks,
the so-called ``acoustic signature''. A large Hubble constant
reduces all peak heights.

As a second remark we note that changes in different
parameters may change certain sections of the spectrum in the same way. 
In particular if one only has access to the primary peak
a very high level of degeneracy will occur. This degeneracy
will then correlate estimates of degenerate parameters, increasing
their marginal variances significantly. Again  $\Omega$ and
$n_S$ are the only two parameters which appear  not
to be much affected by this problem, as their signatures are
sufficiently distinct. For disentangling
$h$ and $\Omega_bh^2$, on the other hand, one should go into the secondary
peak region. In particular the second Doppler peak is essential for
removing the degeneracy between these two parameters. For all other 
parameters one has to rely on rather subtle differences in the spectrum
shape to remove their degeneracy.

These statements are formalized in Fig.~\ref{fig2} where we have
plotted percentage sensitivities (that is $S(l;s_i)s_i^2$) wherever
$s_i\neq 0$, plain sensitivities otherwise. Clearly most of the information
leading to $\Omega$ and the fixing of other parameters which may
have a degenerate effect (like $h$) is contained in the Doppler peaks,
particularly in the secondary Doppler peaks.  The plateau 
carries little relative weight.
The sensitivity $S(l;\Omega)$ has an oscillatory
structure since most of the effects of changing $\Omega$ 
consist of a shift in $l$. Hence around stationary points of the spectrum
there is no information on $\Omega$ whereas the information reaches peaks
near the steepest slopes of the spectrum. Since the shift is multiplicative,
the change in the spectrum is largest at high $l$. Hence the secondary
Doppler peaks' slopes, rather than the main peak summit,  
are the ideal scales for measuring $\Omega$.
For $h$ and $\Omega_b h^2$ most
of the information is in the summit values of the Doppler peaks.
The sensitivities to $n_S$, $r$, and $\tau$ are smoothly distributed all over
the spectrum. Overall, 
the sensitivities for $\Omega$ and $n_S$ are much larger
than for other parameters.

From Fig.~\ref{fig2} one may also draw information on degeneracy.
This may be inferred from similarly shaped  
$S(l;s_i)$ curves. In the absence of noise:
\begin{equation}
{\rm cor}^2(s_i,s_j)={\left(\sum_\ell S(l; s_i)
\sum_\ell S(l; s_j)\right)^{1/2}             
\over \sum_\ell \sqrt {S(l; s_i)S(l; s_j)}}.
\end{equation}
Hence sections of the spectrum where $S(l;s_i)\propto S(l;s_j)$
contribute to a large correlation between the $s_i$ and $s_j$ estimates.
The fact that $S(l;\Omega)$ looks so different from all other curves
then gives credit to the hope that a clean  unambiguous measurement
of $\Omega$ may be achieved by a CMB experiment. 
\begin{figure}
\centerline{\epsfig{file=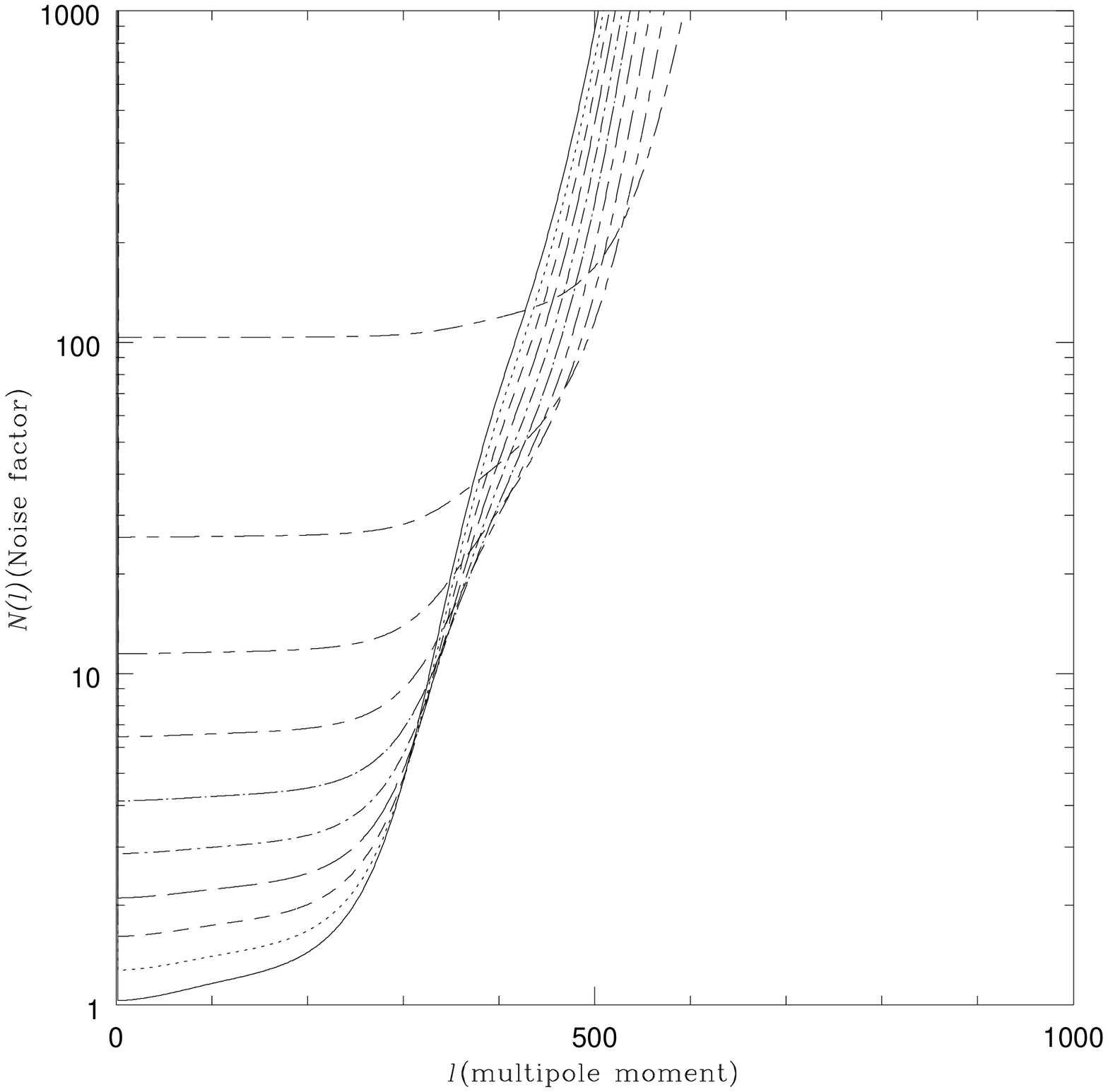,angle=0,width=6cm}
\epsfig{file=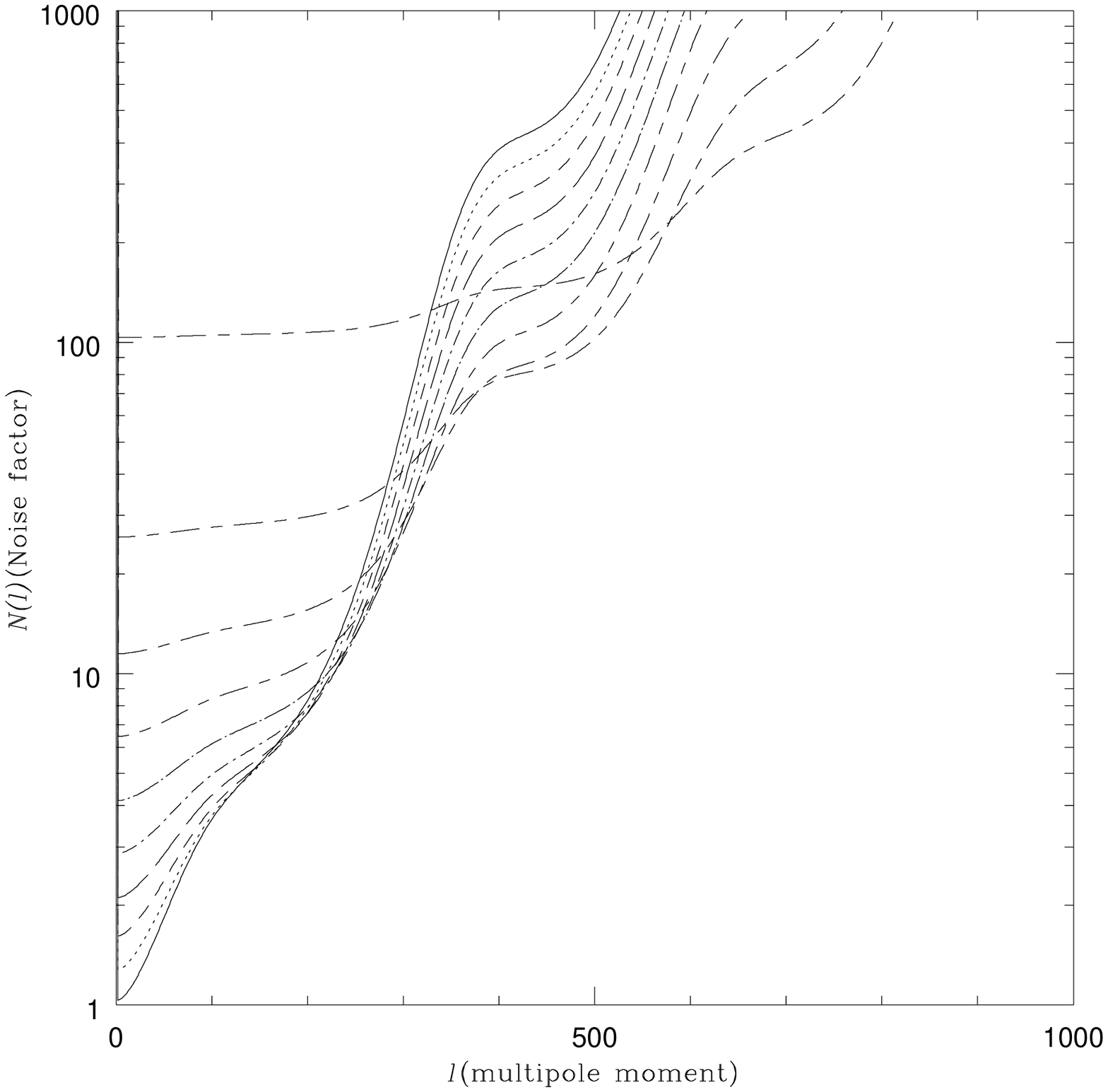,angle=0,width=6cm}}
\caption{The noise factor $N(\ell)$ for sky coverage from $L=200^\circ$
to $L=20^\circ$, for two experimental constraints. On the left
$w^{-1}=(15 \mu K)^2({\rm deg})^2$ and $\theta_b=0.5^\circ$.
On the right $w^{-1}=(60 \mu K)^2({\rm deg})^2$ and $\theta_b=0.2^\circ$.}
\label{fig3}
\end{figure}

\subsection{Noise factors and scanning strategies}

The results of the previous subsection open the problem of the
ideal scanning strategy for measuring cosmological parameters.
Suppose that we have a single-dish experiment with active finite sky-coverage.
For a fixed resolution and noise constraint $w^{-1}$ one now
has to choose the best coverage area.  Best results involve a delicate 
balance between noise and sample variance. Small fields have an increased 
sample variance, and are blind to the low $l$ section of the spectrum. 
However they enable reducing
the effects of noise significantly, allowing for more integration time
to be concentrated on each pixel. A better measurement of high $l$
sections may be then provided. 
The exact balance point between these opposite requirements
depends on the particular question one wants to answer.
For the determination of cosmological parameters this balance point
is set by  demanding that $N(\ell)$ be small where $S(l,s_i)$ is large.
In the previous subsection we showed which scales $l$ should then 
be targeted. 

In Figure~\ref{fig3} we show the effect of reducing the sky-coverage
in two experiments: one with low noise and intermediate resolution,
the other with intermediate noise and high resolution. As the sky coverage
is reduced, the low $l$ section of the spectrum becomes more noisy,
but after a certain cross-over $l$ one actually obtains better estimates. 
In the first experiment the cross over occurs
at $l\approx 400$, when the noise factor has already shot up exponentially.
Therefore in this case it does not seem that reducing sky coverage may be
of any use. The real limitation is resolution and not noise. 
One is better off mapping as much sky as possible in order to reduce 
sampling effects in the $l<300$ sections of the spectrum, where the
experiment fares well. In the
second experiment, on the contrary, by reducing the sky coverage we throw
away the $l<200$ section of the spectrum, but we also greatly improve results
at high $l$. Reducing sky-coverage in this experiment allows for noise
to be defeated effectively, letting the experiment high resolution
manifest itself in useful estimates of $l>400$. Due to the
high sensitivities in $\Omega$ in this section of the spectrum one
may then achieve an accurate measurement of $\Omega$.
The results from these two experiments may then be of comparable
quality. However this quality is achieved by targeting different
sections of the spectrum, and with rather different ideal scanning
strategies.

These two situations are typical. In the first case one relies on a very
accurate mapping of the first peak, by means of low noise, large
sky coverage, and intermediate resolution. In the second, on the contrary,
one makes use of the large sensitivity $\Omega$ has at large $l$,
and targets this section only. Low coverage, and high resolution
are the right combination to do this at intermediate noise. 
\begin{figure}
\centerline{\epsfig{file=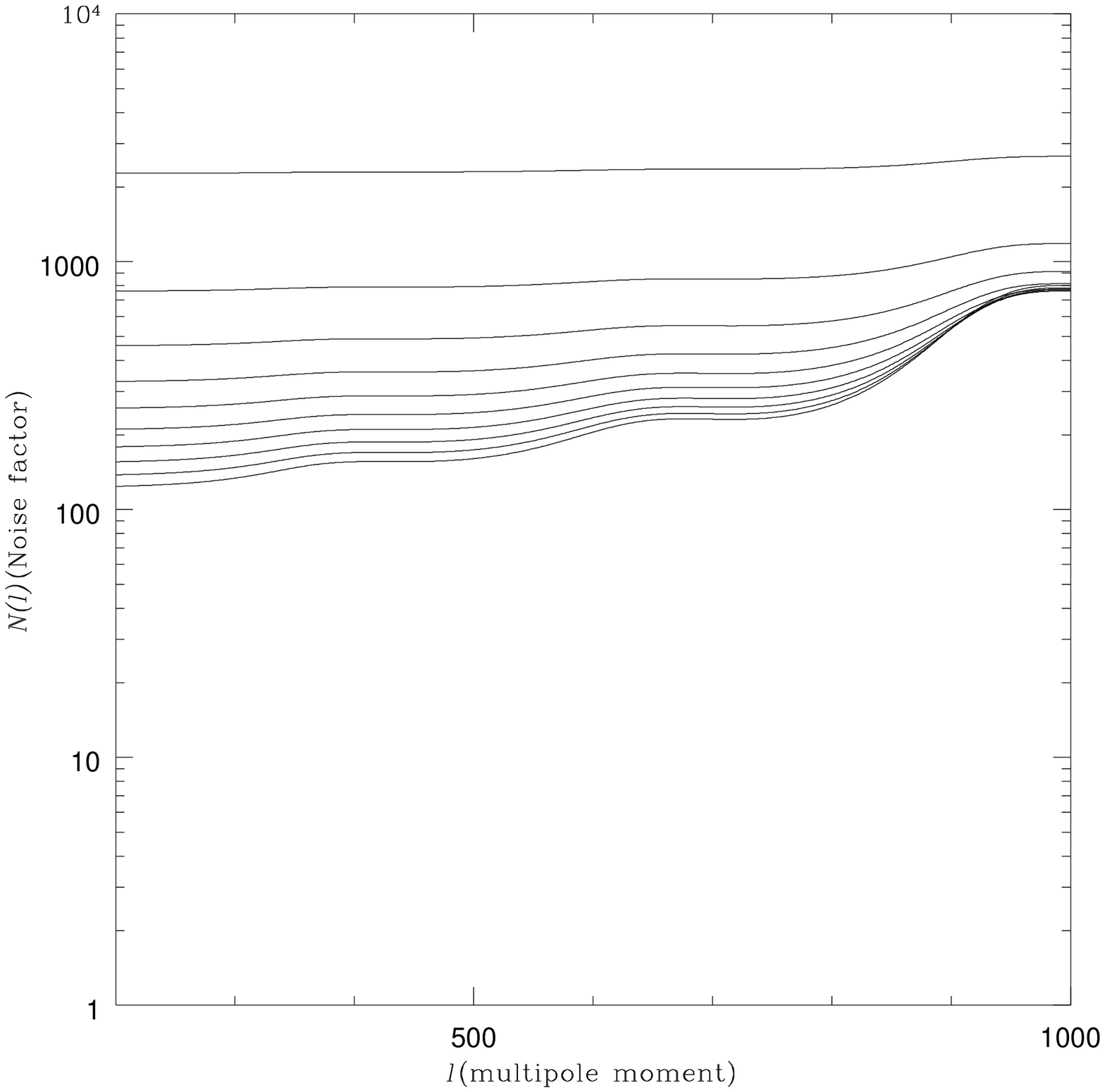,angle=0,width=6cm}
\epsfig{file=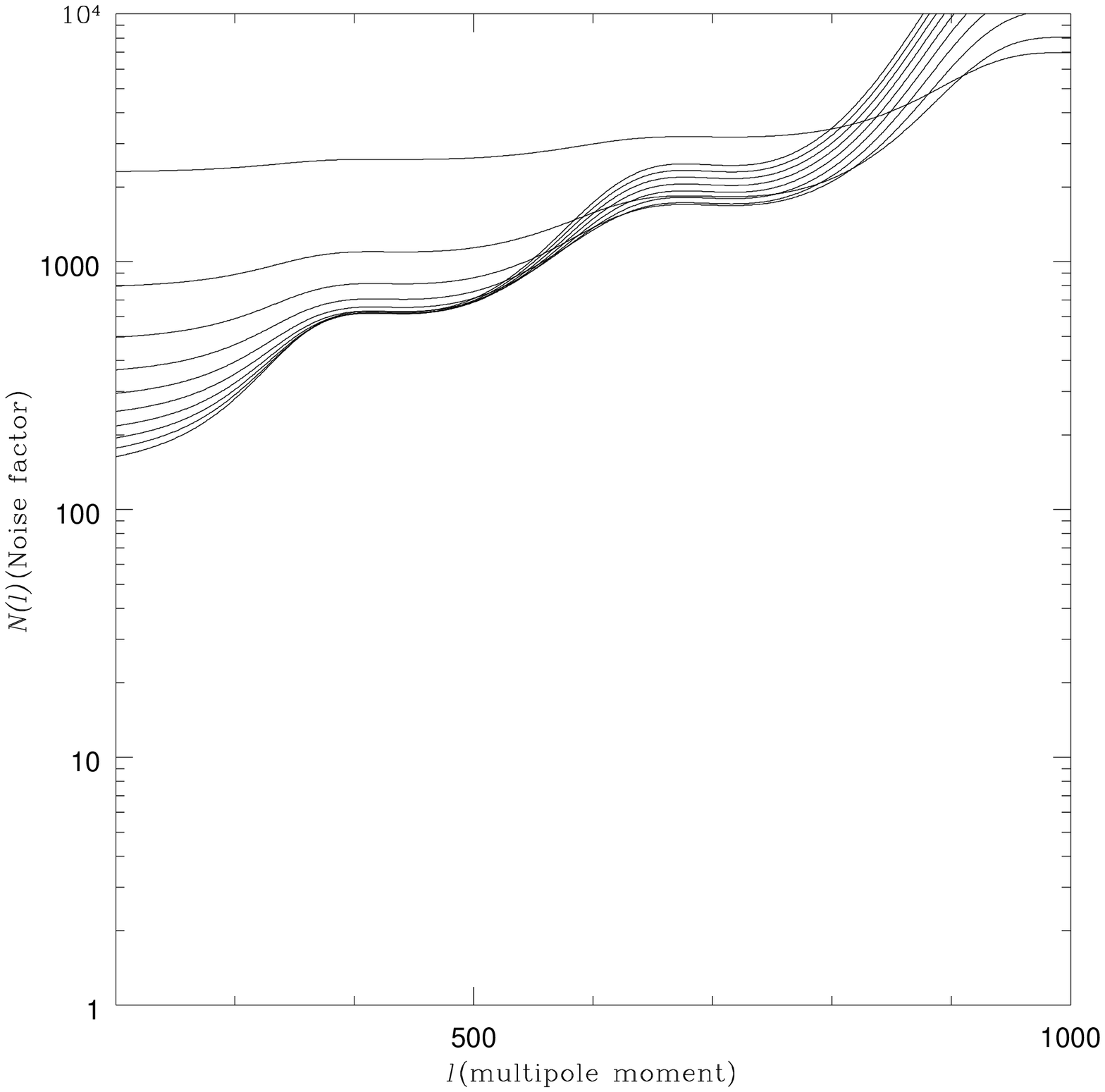,angle=0,width=6cm}}
\caption{$N(\ell)$ for an interferometer
with a primary beam of $\theta_w=4^\circ$, and 
$w^{-1}=(10 \mu K)^2({\rm rad})^{-6}$ (left) and
$w^{-1}=(30 \mu K)^2({\rm rad})^{-6}$ (right). 
The various curves correspond to different numbers of fields
starting at 1 (top curves) up to 19, in steps of two.}
\label{fig4}
\end{figure}

Although one can never give up the uncorrelated mesh approach in 
the case of interferometers, their noise factor may be approximated by 
\begin{equation}
N(\ell)={k_0^2\over \pi n_f}\left(1+{w^{-1}\Omega^{s2} n_f\over 
C(\ell_i)}\right)^2.
\end{equation}
In Figure~\ref{fig4} we have plotted  $N(\ell)$ for an interferometer
with a primary beam of $\theta_w=4^\circ$, and 
$w^{-1}=(10 \mu K)^2({\rm rad})^{-6}$ (left) and
$w^{-1}=(30 \mu K)^2({\rm rad})^{-6}$ (right). The various curves correspond
to different numbers of fields.  The most noticeable feature
is that although the overall noises are high, they do not increase by
much with $\ell$. As a result interferometers provide the lowest 
$N(\ell)$ at high $\ell$. The main limitation is clearly sample
variance, especially if the noise is of order 
$w^{-1}=(10 \mu K)^2({\rm rad})^{-6}$. For this noise level the best
scanning strategy will always be to try and map as many fields as
possible, granted the limitations of parallel point source subtraction.
Sample variance becomes less relevant at very high $\ell$ however,
and so even with a limited number of fields one may fare well at high
$\ell$ with interferometers. Once again, given the high sensitivity
of the spectrum to $\Omega$ at high $\ell$ one may expect a good
measurement of $\Omega$ with an interferometer. Hence interferometers
may be expected to be competitive with the previous two experimental 
strategies, for yet another reason, and with yet another ideal scanning
strategy.

\section{Ideal CMB experiments for weighing the Universe near sCDM}
\label{cdm}

The heuristic arguments of the previous Section will now be replaced
by a full calculation of errors in $\Omega$ for a large section of
experiment parameter space. As stated before
best results in any CMB experiment involve a delicate 
balance between noise and sample variance. 
The exact balance point depends on the particular 
question one wants to answer.
In \cite{us} we considered the problem of detecting secondary Doppler
peaks. In comparison, we have found that the problem of the measurement
of $\Omega$ marginalizing with respect to a large set of paramaters
is more sensitive to sample variance, and therefore always
requires a larger optimal coverage area. If the set of uncertain parameters
is reduced, however, not only do errors go down sharply, but also
sample variance again becomes of secondary importance. We will illustrate 
this point in Section~\ref{fewpar}.
Unless otherwise stated, we will, in this section, marginalise over
all the parameters  ${\bf s}=\{Q,\Omega,$ $\Omega_b h^2,$ $h,n_S,\Lambda,$
$r,n_T,\tau,\Omega_{\nu}\}$, except for $\Omega_{\nu}$ which we assume
to be zero, i.e. we assume no massive neutrinos.

We will highlight three corners of experiment parameter space which 
we found the most promising: satellite experiments, ground based 
deep-patch experiments, and ground based interferometers. 

\subsection{Satellite experiments}

An instrumental noise level of $w^{-1}=(7.5 \mu K)^2({\rm deg})^2$
is often quoted as realistic for a satellite experiment (\cite{kn,j,j1}).
Some of the COBRAS-SAMBA channels achieve a factor of 5 improvement on
this \cite{cobras/samba}. Whatever the case one must bear in mind that these
instrumental noises are then considerably deteriorated by the need
to deconvolve foregrounds. For definiteness we shall here consider
the following two noise levels as renormalized after foreground subtraction:
$w^{-1}=(15 \mu K)^2({\rm deg})^2$ and $w^{-1}=(30 \mu K)^2({\rm deg})^2$.
We believe that either figure is realistic, but the first corresponds
to an optimistic hope, the latter to a pessimistic expectation.
A remark on normalization should be made. If in future experiments
$Q$ comes out very different from COBE's $Q=18\mu K$, 
then this has the effect of renormalizing
$w^{-1}$ by the same factor.  Hence a lower normalization would
boost $w^{-1}$ unpleasantly.

\begin{figure}
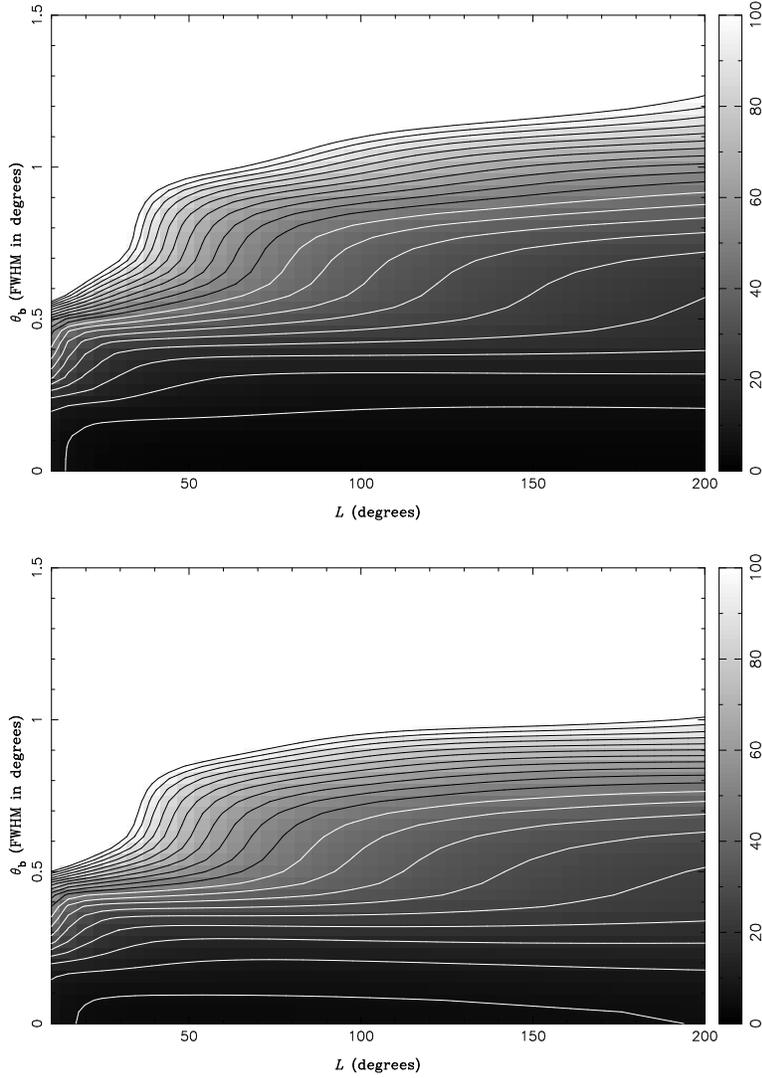

\centerline{\epsfig{
file=scdm_sd15.ps,angle=-90,width=10cm}}
\qquad
\centerline{\epsfig{
file=scdm_sd30.ps,angle=-90,width=10cm}}
\caption{Percentage standard errors in $\Omega$ for $w^{-1}=(15 \mu K)^2({\rm
deg})^2$ (top) and $w^{-1}=(30 \mu K)^2 ({\rm deg})^2$ (bottom). In both
cases, isolines are plotted from 5\% to 100\% at 5\% increments.}
\label{fig5}
\end{figure}

We have plotted the percentage errors in $\Omega$ for these two cases
in Fig.~\ref{fig5}. 
Clearly, even considering active limited sky 
coverage, the ideal coverage area is all-sky for the $w^{-1}=(15 \mu K)^2({\rm
deg})^2$ case, and also for $w^{-1}=(30 \mu K)^2({\rm deg})^2$, except
perhaps for very small beamwidths $\theta_b \leq 0.1^\circ$.
Measuring $\Omega$ as one of a large set of unknown parameters
is obviously very sensitive to sample-variance. Even for noise
levels that are not too low, like the ones considered, we are dominated by
the need to reduce sample variance, not noise. 

As explained further in Section~\ref{fewpar} there are ambiguities
in the covariance matrix approach which allow for some
uncertainty in the errors. Thus, in the calculations presented here, we
present the most pessimistic results. This resulting error
in $\Omega$ is of the order of 20\% for 
$w^{-1}=(15 \mu K)^2({\rm deg})^2$ and
$\theta_b=0.5^\circ$. These errors are quickly reduced to about
$5-10\%$ with an improved resolution of $\theta_b=0.2^\circ$.
Therefore a satellite experiment is always ideally scanned with the largest
possible sky coverage, and the final result benefits enormously from
improving the resolution. 

\subsection{An intermediate noise deep-patch experiment}

\begin{figure}
\centerline{\epsfig{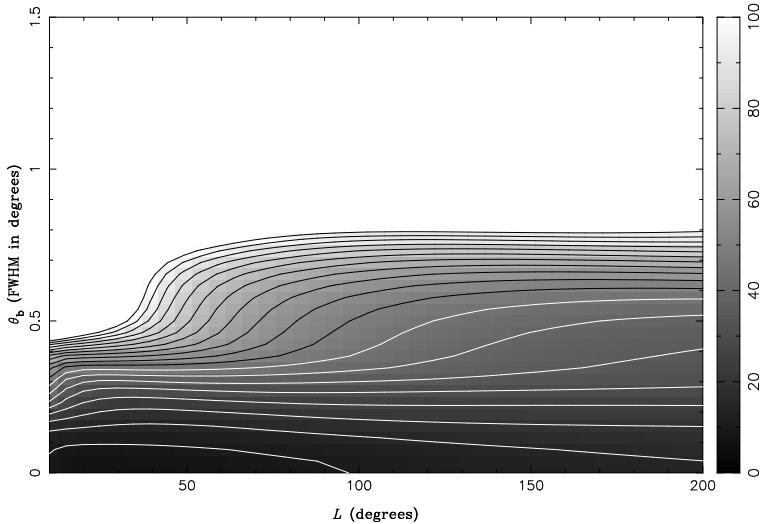}}
\caption{Percentage standard errors in $\Omega$ for 
$w^{-1}=(60 \mu K)^2({\rm deg})^2$. Isolines are plotted 
from 10\% to 100\% in 5\% increments.}
\label{fig6}
\end{figure}

If one is observing from the ground, the noise level is always larger.
This is because the atmosphere, as well as ground
spill-over, acts so as to deteriorate the experiments sensitivity, often
by a large factor. 
If, however, one takes advantage of the fact that a large aperture is
no longer a problem, then one may make up for this with increased resolution.
The ideal coverage area is now very far from all-sky, but, in any case, a small
patch of the sky is more feasible from the ground.
We illustrate this promising corner of experiment parameter space in
Fig.~\ref{fig6}. We have assumed $w^{-1}=(60 \mu K)^2({\rm deg})^2$.
A error in $\Omega$ of around 10 percent may then be achieved
with a resolution around $0.1^\circ$, with a coverage area corresponding
to $L=20-40^\circ$. It is clear from the figure that, for such an experiment,
increasing the coverage area is extremely undesirable.

\subsection{An intermediate noise interferometer experiment}

\begin{figure}
\centerline{\epsfig{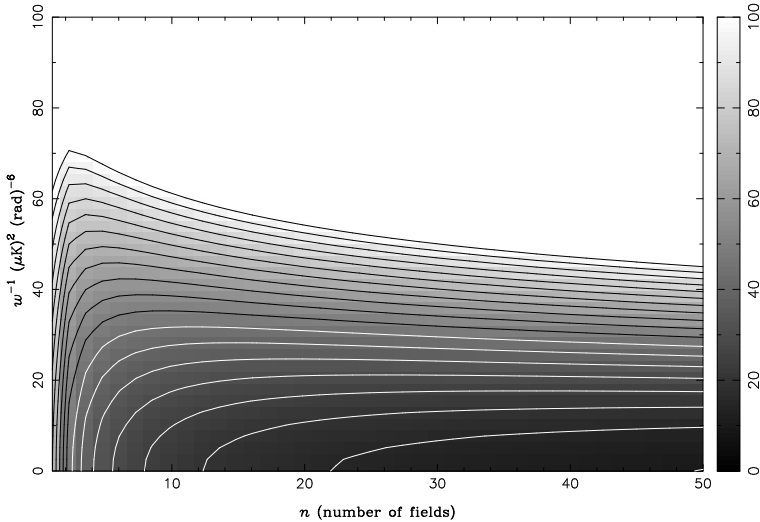}}
\caption{Percentage standard errors in $\Omega$ for an interferometer
with $w^{-1}$ up to $(100 \mu K)^2({\rm rad})^{-6}$. 
We have assumed a primary beam of $\theta_w=4^\circ$.
Isolines are plotted from 15\% to 100\% in 5\% increments.}
\label{fig7}
\end{figure}

In Fig.~\ref{fig7} we show the percentage error in $\Omega$ for
an interferometer with a primary beam of $\theta_w=4^\circ$.
We have assumed that a uniform $uv$-plane coverage up to $\ell=1000$
has been achieved by means of a sufficient number of elements, and 
their arrangement in a suitable geometry. With these assumptions we
have then allowed the noise parameter 
$w^{-1}$ as defined for interferometers
to vary between zero and $(100 \mu K)^2({\rm rad})^{-6}$.
Finite sky-coverage errors due to small size and 
limited number of fields are now substantial. The sample variance
in the estimates of $\Omega$ for the interferometer considered 
can be read off from the $w^{-1}=0$ axis. Although these errors
are not small, their limiting values may be approached
even for reasonably large values of $w^{-1}$. The need for point-source
subtraction typically limits the number of fields, so unfortunately
one may not be able to obtain the best result for a given level of 
noise $w^{-1}$.

We would expect the next generation of CMB interferometer experiments,
such as the VSA, VCA and CBI to attain noise level of below
$w^{-1}\approx (10 \mu K)^2({\rm rad})^{-6}$, for the $uv$-coverage
assumed above, as compared to 
$w^{-1}\approx(200 \mu K)^2({\rm rad})^{-6}$
for the CAT experiment, after foreground subtraction.
Assuming a fixed observation time of about 2 years, one could hope to determine
$\Omega$ to within about 15--20 percent by observing roughly 20 fields.

\section{Dependence on number of undetermined parameters}
\label{fewpar}

As one might expected, the number of unknown parameters in a given model
has a profound effect on the overall error in determining $\Omega$ 
(or any other parameter). In this Section we illustrate 
this important effect, and also highlight
some weaknesses in this method of analysis used, particularly in relation
to the computation of the curvature matrix. 

\subsection{Margins of error in covariance matrix estimates}

\begin{figure}
\centerline{\epsfig{file=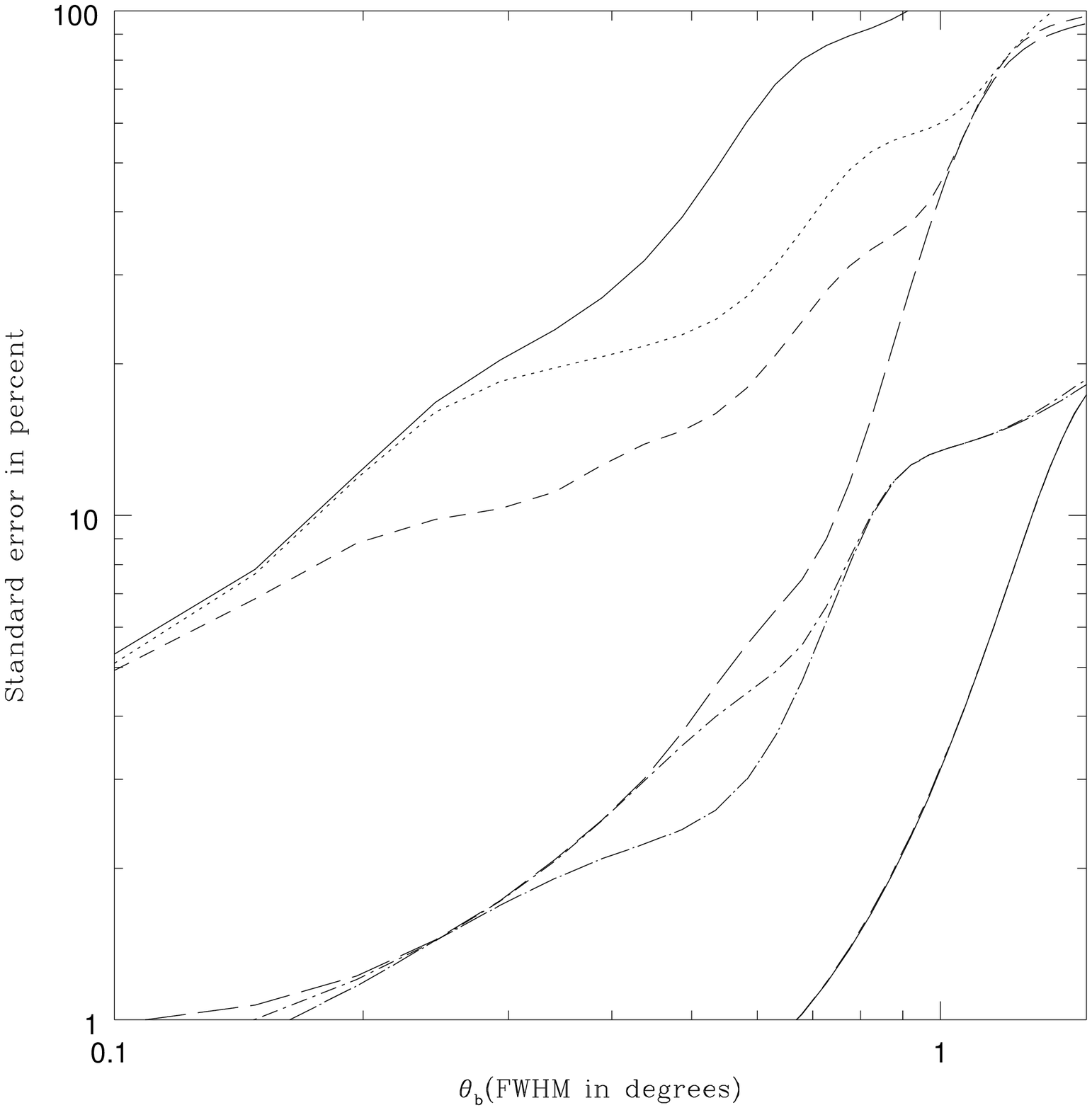,angle=0,width=6cm}
\epsfig{file=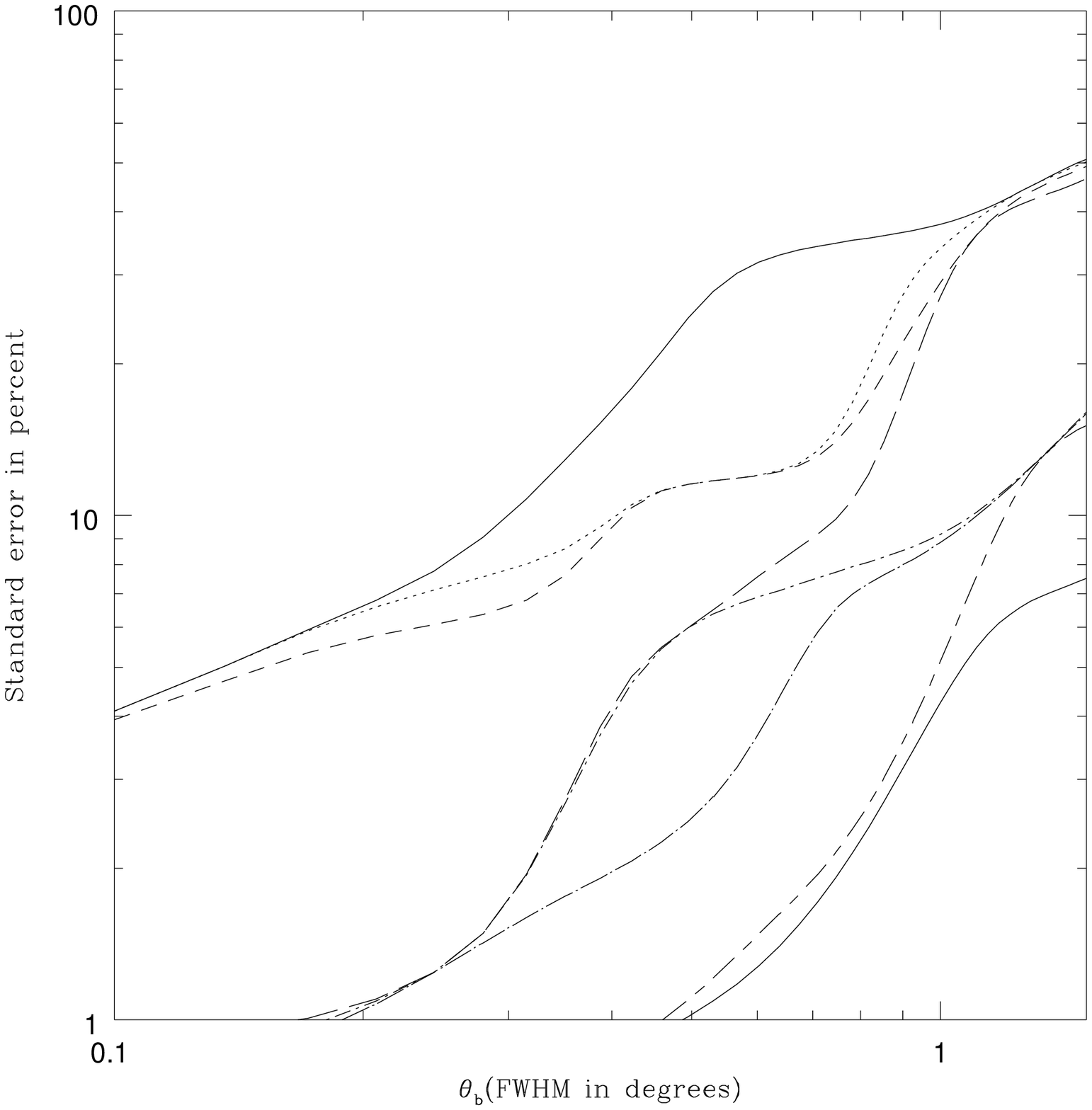,angle=0,width=6cm}}
\caption{Percentage standard errors in $\Omega$ for an all-sky 
experiment with $w^{-1}=(15 \mu K)^2({\rm deg})^2$, as one adds
more parameters into the problem. The lowest curves consider only
$(Q,\Omega)$ to which parameters are then added in the same order
as in the full vector ${\bf s}=\{Q,\Omega,$ $\Omega_b h^2,$ $h,n_S,\Lambda,$
$r,n_T,\tau,\Omega_{\nu}\}$. }
\label{fig8}
\end{figure}

The covariance matrix formalism suffers from an ambiguity in the definition
of the derivatives $\partial C_\ell/\partial s_i$. These have to be computed
by means of finite differences, which can never realistically be very small.
A particularly difficult derivative is $\partial C_\ell/\partial \Omega$.
This may be inferred from the approximate formula, valid 
in the vicinity of sCDM
\begin{equation}
C(\ell,\Omega,h,\Omega_bh^2)\approx
C(\ell\Omega^{1/2},1,h\Omega^{1/2},\Omega_bh^2),
\end{equation}
resulting in the relation
\begin{equation}\label{implicit}
{\partial C_\ell\over \partial \Omega}{\Big {|}}_{\Omega=1}=
{\ell\over 2}{\partial C_\ell\over \partial \ell}
+{h\over 2}{\partial C_\ell\over \partial h}
\end{equation}
The effects of $\Omega$ on the plateau can be fitted by a multiplicative
factor \cite{j1} which has zero derivative at $\Omega=1$, that is,
it is a saddle point. Hence the low $l$ signatures of low $\Omega$ models
do not appear at all in the linear approximation, in the vicinity
of sCDM. Making use of the prescription (\ref{implicit}) always 
results in larger errors
than say simply computing the derivative with $\Delta\Omega=.05$.
This is not at all surprising, as even such a small difference in 
$\Omega$ results in a difference in $\ell$ of the order $\Delta\ell
\approx\ell\Delta\Omega/2$. This corresponds to taking a finite difference
$\Delta\ell$ in (\ref{implicit}) which varies with $\ell$ and is always
much larger than 1 at large $\ell$. In general increasing the size
of the differences used in the computation of the derivatives has the effect
of reducing the errors. The reason we bring up this numerical 
problem is that it may have a physical implication.  If the errors
are of the order of $10\%$ then it may make more sense to use large finite
differences, rather than carefully computed derivatives.
Therefore the results presented in the previous sections (where we
used very fine grids  and also (\ref{implicit})) 
are to be seen as conservative
estimates. A non-linear analysis will always provide smaller errors.

In Fig.~\ref{fig8} we showed the percentage errors in $\Omega$ obtained
by two prescriptions for computing the derivatives. On the left
we have used the implicit formula (\ref{implicit}), on the right
we have simply computed a finite difference in $\Omega$. We have also
plotted the effects of adding more and more parameters into the problem.
The lower curves invert only $\{Q,\Omega\}$. We then add more and more 
parameters, in the order in which they appear in ${\bf s}$.
This picture illustrates the uncertainties inherent in this formalism.
It also shows how the errors may change dramatically by adding more and 
more parameters into the analysis. If one neglects massive neutrinos
and uses a finite difference prescription to compute the curvature matrix
then an error of 10\% in the measurement of $\Omega$ is obtained for
$w^{-1}=(15 \mu K)^2({\rm deg})^2$ and $\theta_b=0.5^{\circ}$ 
(as in \cite{j,j1}). By considering one massive neutrino this
error goes up to 20\% with the same prescription. If instead
of using finite differences one computes $\partial C_\ell/\partial \Omega$
implicitly, the error goes up to 30\%.

\subsection{The sample-variance/noise balance and undetermined parameters} 

\begin{figure}
\centerline{\epsfig{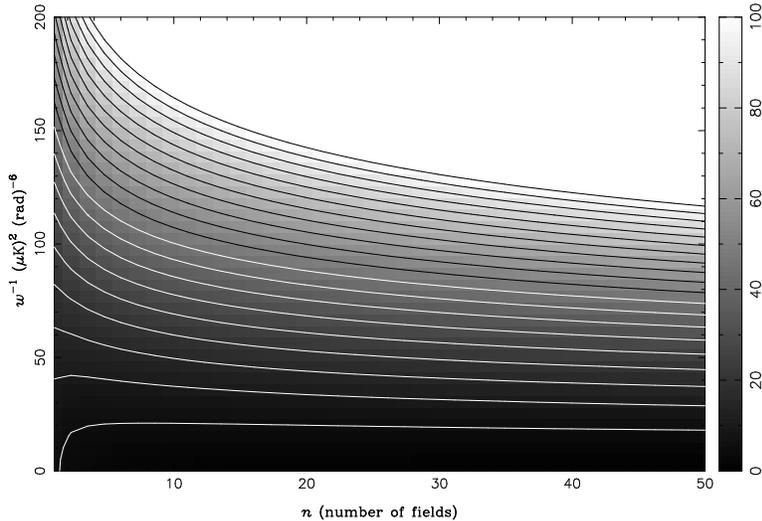}}
\caption{Percentage standard errors in $\Omega$ for an interferometer
with $\theta_w=4^\circ$, considering only the 4 parameters
$\{Q,\Omega,$ $\Omega_b h^2,h\}$. Isolines are plotted from 5\% to
100\% in increments of 5\%.}
\label{fig9}
\end{figure}

If one assumes the values of some of the parameters, so that fewer of
them are allowed to vary, then sample variance
naturally becomes less important. In this case, for example,
it is no longer necessary to observe such a large number of fields 
to obtain a successful interferometric 
determination of $\Omega$. We may illustrate this point by
considering the interferometer experiment discussed in the previous
section, but this time fixing the values of all but four of the
model parameters, and working only with the set
$\{Q,\Omega,$ $\Omega_b h^2,h\}$.  As shown in Figure~\ref{fig9},
with less than 10 fields one could now obtain a better than 5\%
determination of $\Omega$ for $w^{-1}\approx 
(10 \mu K)^2({\rm rad})^{-6}$. Moreover, it is clear that
for relatively high noise levels,
one could concentrate on a single field, to overcome the instrumental
noise, and still obtain a useful result; this is because
the sample variance in a single field in this context is only
$5-10\%$. Thus, even for
$w^{-1}\approx (50 \mu K)^2({\rm rad})^{-6}$, and for a single field,
one could now measure $\Omega$ with an error of about $10-15\%$.

\section{Interferometers and other cosmological parameters}
\label{intsagain}

Although not central to our discussion, it is interesting to explore
the marginal variances for cosmological parameters other than
$\Omega$, in particular for interferometers. This is because interferometers
may be expected to provide the least noisy estimates
of the power spectrum at high $\ell$. Hence the detection of subtle but
valuable high $\ell$ signals, such as the acoustic oscillation signature, 
are obvious goals for such experiments.
\begin{figure}
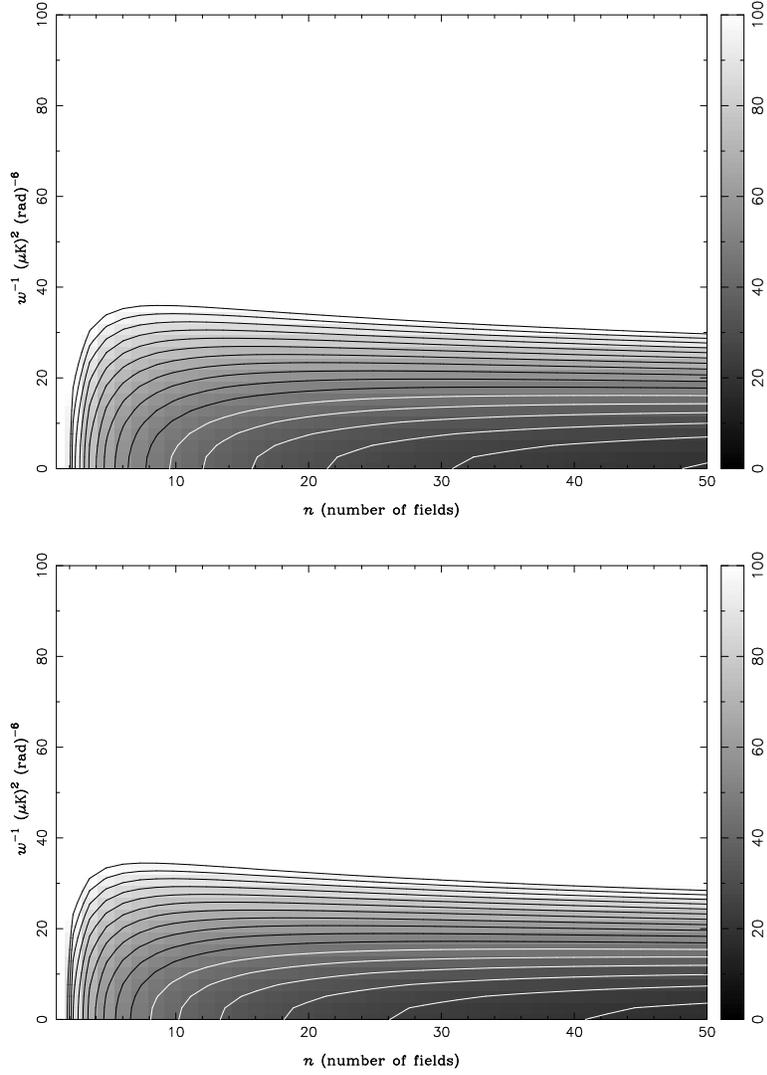

\centerline{\epsfig{
file=scdm_int4_obh2.ps,angle=-90,width=10cm}}
\qquad
\centerline{\epsfig{
file=scdm_int4_h.ps,angle=-90,width=10cm}}
\caption{Percentage standard errors in $\Omega_bh^2$ (top) and
$h$ (bottom) for an interferometer with $\theta_w=4^\circ$, allowing
all parameters to vary (but assuming no massive neutrinos). Isolines
are plotted from 20\% to 100\% in increments of 5\%.}
\label{fig10}
\end{figure}

In Fig.~\ref{fig10} we show the percentage errors in the parameters
$\Omega_bh^2$ (top) and $h$ (bottom), for an interferometer with a
primary beam width of $4^\circ$,  and for which we have allowed all
the CDM parameters to vary (but assuming no massive neutrinos).
Estimates of the two parameters illustrated in the figure are derived mainly
from the relative positions and height of the Doppler peaks in the CMB
power spectrum, and it is clear from the figure that a reasonable
determination of these parameters is possible. Assuming a noise
level of $w^{-1}\approx (10 \mu K)^2({\rm rad})^{-6}$, the error
in both $\Omega_bh^2$ and $h$ is about 35\%, even assuming such
a large parameter set. As discussed above,
however, these errors can be greatly reduced if the values
of some of the other parameters, such as $n_s$ or $\Lambda$, 
can be fixed by other means.

\section{Ideal scales away from the vicinity of sCDM}\label{ncdm}

So far we have concentrated our attention on the errors associated
in estimating $\Omega$ (and other parameters), assuming the sCDM
values for the model parameters. In this section, we conclude by
making a brief investigation of how the errors in the
determination of  $\Omega$
may be much larger if the universe is not described by the sCDM model.

In particular, we will consider two non-standard CDM models
which illustrate two ways in which the problem of measuring $\Omega$ 
may be more difficult. The first is an
open universe with $\Omega=0.3$, which has all other model parameters
set to their sCDM values; we call this model oCDM. If the signal
is of this type then the ideal scales for weighing the Universe
will be at much larger $\ell$. Consequently more stringent bounds
will be put on the resolution of single-dish experiments.
Interferometers, however, will be little affected.

Our second model
has $\Omega=1$, but instead we assume a scalar tilt of $n_S=0.8$.
The remaining parameters all
have their sCDM values, and we call this model tCDM.
For this type of model the intensity of the signal at the Doppler
peaks is much smaller, requiring larger sensitivities from both
inteferometers and single-dish experiments. This obstacle
for an accurate measurement of $\Omega$
occurs also for signals coming
from theories with low $Q$, non-zero $r$ (a gravitational
wave component), or non-zero $\tau$ (late reionization).

\subsection{Satellite experiments}

In Fig.~\ref{fig13} we plot the percentage errors in $\Omega$ for
oCDM (top) and tCDM (bottom), for a satellite experiment with
$w^{-1}=(30 \mu K)^2({\rm deg})^2$. 
\begin{figure}
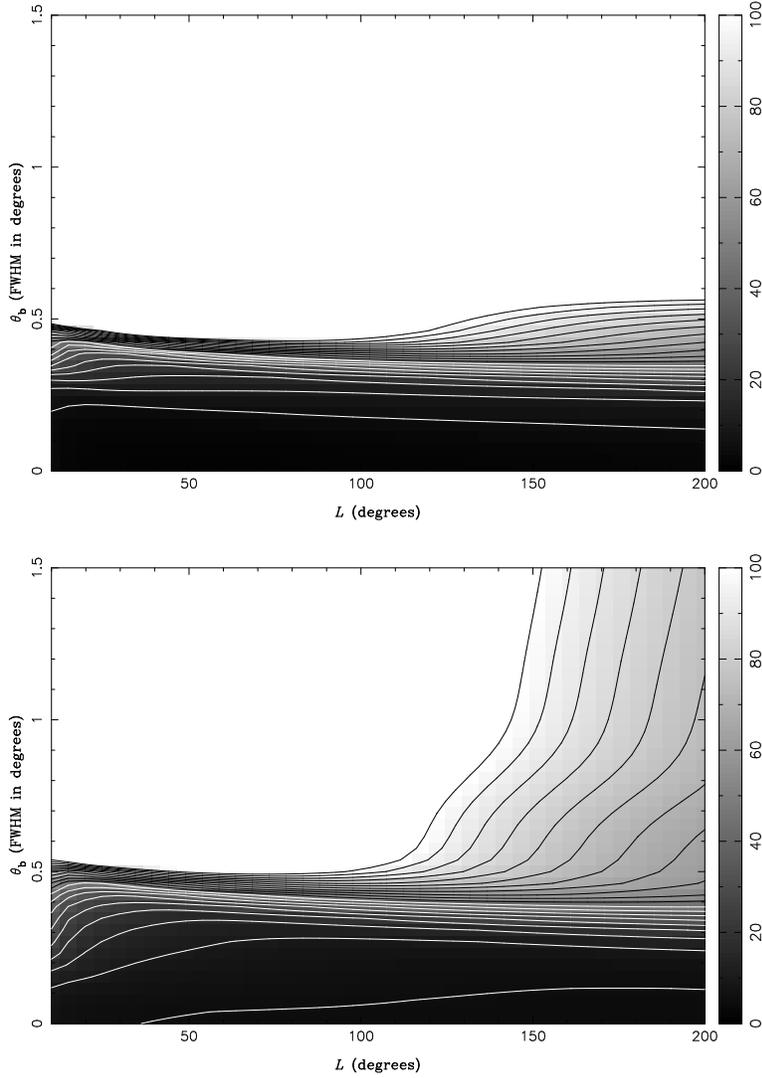

\centerline{\epsfig{
file=ocdm_sd30.ps,angle=-90,width=10cm}}
\qquad
\centerline{\epsfig{
file=tcdm_sd30.ps,angle=-90,width=10cm}}
\caption{Percentage standard errors in $\Omega$ for a satellite
experiment with $w^{-1} = (30 \mu K)^2({\rm deg})^{2}$ in a universe
with all parameters equal to those of sCDM, but with $\Omega=0.3$ (top) and
$n_S=0.8$ (bottom). Isolines are plotted from 5\% to 100\% (top) and
from 10\% to 100 \% (bottom), in increments of 5\%.}
\label{fig12}
\end{figure}
It is clear from the figure
that for oCDM we require a resolution of better than $0.4^\circ$ in
order to determine $\Omega$ to any reasonable accuracy. 
Overall in comparison with sCDM
the error plot gets squashed towards lower $\theta_b$.
This is what one
would expect, since the Doppler peaks in such a
model are shifted to significantly higher multipoles, and so a finer
angular resolution is required to determine their position. For
sufficiently fine resolution and with this relatively low level of
noise, we also see that all-sky coverage is now not a great advantage.
This is because 
the sample variance on the CMB power spectrum at the position of the 
Doppler peaks is reduced by their shift to higher multipoles.
In any case, however, for a resolution of $0.1^\circ$, one could hope
to achieve an accuracy of better than 5\%, with little dependence on
the area of sky covered.

For tCDM, by comparing the bottom panels of
Figs~\ref{fig5} and \ref{fig12}, we see that again the errors
are much larger. Overall the effect of reducing the
power in the Doppler peaks by considering, say, a theory
with $n_S<1$, is similar to increasing the noise parameter
$w^{-1}$. For this reason all-sky coverage is now far from ideal.
A 10\% determination of $\Omega$ under these conditions would
now require a resolution of about $\theta_b\approx 0.2^{\circ}$
with an ideal scanning area corresponding to $L\approx 100^{\circ}$.

\subsection{Interferometer experiments}

\begin{figure}
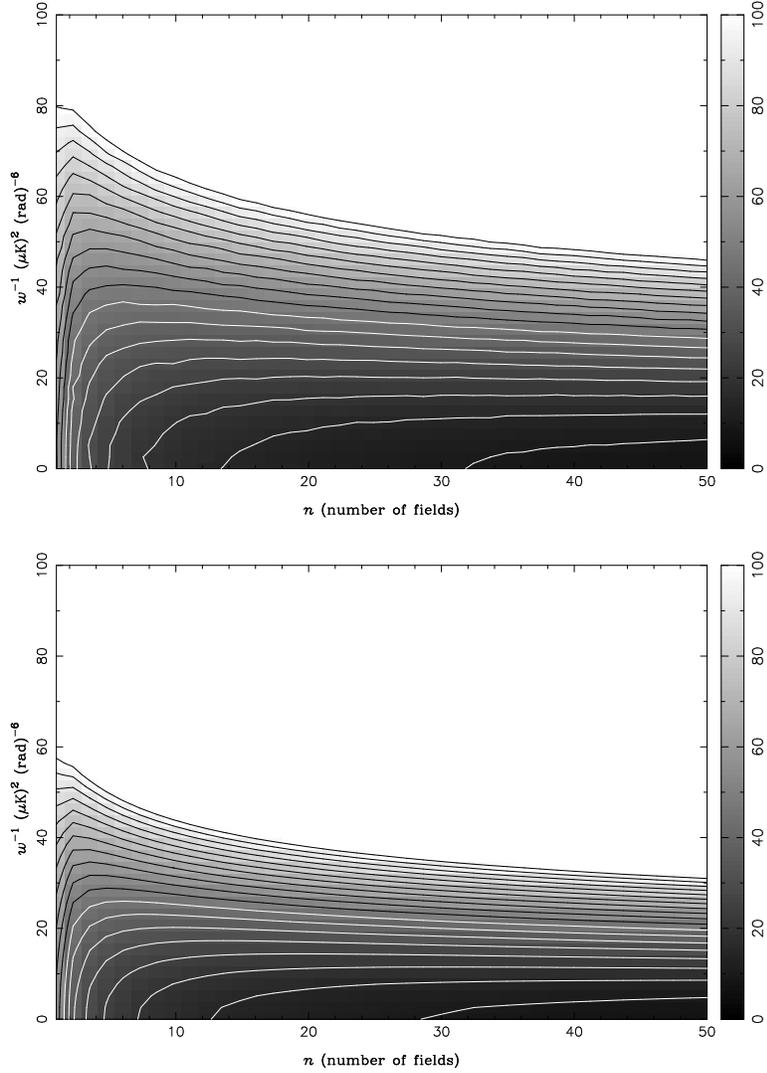

\centerline{\epsfig{
file=ocdm_int4_ot.ps,angle=-90,width=10cm}}
\qquad
\centerline{\epsfig{
file=tcdm_int4_ot.ps,angle=-90,width=10cm}}
\caption{Percentage standard errors in $\Omega$ for the interferometer
experiment with $\theta_w=4^\circ$, for a model
with all parameters equal to those of sCDM, but with $\Omega=0.3$ (top) and
$n_S=0.8$ (bottom). Isolines are plotted from 10\% to 100\% (top) and
10\% to 100\% (bottom) in increments of 5\%.}
\label{fig13}
\end{figure}

We have also investigated the performance of interferometer
experiments in these two non-standard cosmologies, and the results are
given in Fig.~\ref{fig13}.

From the figure we see that interferometers suffer far less in oCDM
than satellite experiments. In fact, there is very little change in
the accuracy to which $\Omega$ can be measured.  If anything
the situation improves wherever interferometers are limited
by their large sample variance.
This is not
surprising, since the $uv$-coverage is assumed constant up to 
$\ell \approx 1000$. This still easily includes the relevant 
Doppler peaks in this model. Therefore the noise factor in
the sensitive sections of the spectrum does not increase
in the same way as it does for a single dish experiment. 
As an example, for $w^{-1}\approx (50 \mu K)^2({\rm rad})^{-6}$, 
we can expect an
accuracy of about 15--20\% by observing 10--20 fields.

In the tCDM model, interferometers are affected in the same way
as single-dish experiments. Again one suffers from a weaker
Doppler peak signal, effectivelly similar to a larger noise
parameter. Nevertheless, 
for $w^{-1}\approx (10 \mu K)^2({\rm rad})^{-6}$ with 
20 observed fields, the error in $\Omega$ is still around 20\%.

\section{Summary and concluding remarks} 
In this paper we reused the linear approximation tools developed
in \cite{j,j1} in a broader context. We considered a larger class of 
experiments, more evasive signals, and formalism implementations
producing less spectacular results. We then set off looking
for the ideal experiment for weighing the Universe. The
exercise is clearly somewhat naive. Political forces are
more likely to determine experimental design than scientific reasons.
Nevertheless it is interesting to note from our plots
that accurate measurements of $\Omega$ may be expected even from 
poorly designed experiments.

In a rough summary we found three corners in experiment parameter
space which, for different reasons, and using different strategies,
provide estimates for $\Omega$ with useful errorbars.  Satellite
experiments make use of their low noise and intermediate resolution to
map the first peak (and maybe part of the second) very accurately.
They rely on beating sample variance by means of a large sky coverage.
Ground-based and balloon borne experiments may make up for their
larger noises with improved resolutions.  This results in a better
mapping of the second and maybe third Doppler peak.  Small-sky
coverage is used as a strategy to beat noise.  The resulting increase
in the sample variance is expected to be less of a problem given the
larger $\ell$ at which the relevant measured features
occur.  Both these experimental strategies suffer dramatically if the
Doppler peaks happen to be at larger $\ell$ than in the standard
model, such as is the case with open models.

Interferometers constitute a third viable avenue.
They may be expected to map the first peak very poorly, but they will do
outstandingly well in mapping the secondary peaks,
even when compared with high-resolution single-dish
experiments. Again a small coverage area technique
is employed so as to reduce the effects of noise.
The sample variance is always large, but again one may hope
that its practical effects are small due to the
high $\ell$ features which should be targeted.
However, this is not always the case, and in fact
interferometers are often limited by sample variance.
For this reason experimental interferometric parameters exist, for which 
performance is improved if the relevant features are at
higher $\ell$, as with open models. All this seems to indicate
that one should explore more than one experimental avenue
in order to avoid unpleasant surprises.

We conclude with a couple of remarks on how to improve the
linear analysis performed here. Clearly one should compute
the likelihood directly from the $C_\ell(s_i)$ function.
This is feasible but computationally demanding. Typically
the non-linear analysis tends to predict smaller errors.
Also one should go beyond the uniform prior assumption 
used here. A variety of information on all of the $s_i$
parameters exists, coming from all sorts of other sources.
This should be incorporated in the priors, naturally reducing
the errors of the final estimates. For instance we have found
errors in $\Omega_\nu$ of order 0.9. These clearly are not 
compatible with what we already know about the mass of the neutrinos,
should there be a heavy neutrino. The same can be said for
some of the errors found for $\Omega_b$, $h$, $n_S$. 
The final analysis waiting to be done should therefore not
only go beyond the linear approximation, but also proceed
to set up priors incorporating the vast amount of information
coupled to the parameter determination problem posed by the CMB.

In closing we should mention that marginal variances may be perhaps
somewhat misleading. We have used them simply because they
allow for a nicer display of  results. However errorbars 
are in fact ellipsoids in the parameter space $\{s_i\}$.
In the non-linear approximation they become rather irregular
bubbles of maximum likelihood. Such bubbles cannot be visualized
but they encode all the information one may extract from a given
experiment. Their computation and storage may turn out to 
be computationally very demanding, but such is the task one should
perform when data finally starts pouring in.

\section*{Acknowledgements}
We would like to thank Marc Kamionkowski for helpful tips at the
start of this project. We also thank Uros Seljak and Matias
Zaldarriaga for providing us with their Boltzmann code \cite{uros}, which 
we used in all our calculations.
We acknowledge  St.John's College (J.M.), and Trinity Hall (M.H.), 
Cambridge, for support in the form of research fellowships.
J.M. thanks the Royal Society for a University Research Fellowship.
J.M. also wishes to thank Kim Baskerville for reading the manuscript,
and to acknowledge MRAO and DAMTP (Cambridge)
as his host institutions for most of the period when this work was done.

\end{document}